\documentclass[aps,prb,twocolumn,showpacs,amsmath,amssymb]{revtex4}
\usepackage{graphicx}
\usepackage{dcolumn}
\usepackage{bm}

\begin{document}
\title{Weak ferromagnetism in Mn nanochains on the CuN surface}
\author{A.N. Rudenko$^{1,2}$, V.V. Mazurenko$^{1,2}$, V.I. Anisimov$^{1,3}$ and A.I. Lichtenstein$^{2}$}
\affiliation{$^{1}$Theoretical Physics and Applied Mathematics Department, Urals State Technical University, Mira Street 19,  620002
Ekaterinburg, Russia \\
$^{2}$Institute of Theoretical Physics, University of Hamburg, Jungiusstrasse 9, 20355 Hamburg, Germany  \\
$^{3}$Institute of Metal Physics, Russian Academy of Sciences, 620219 Ekaterinburg GSP-170, Russia}
\date{\today}

\begin{abstract}
We investigate electronic and magnetic structures of the Mn chains supported on the CuN surface using first-principle LSDA and LDA+U calculations. 
The isotropic exchange integrals and anisotropic 
Dzyaloshinskii-Moriya interactions between Mn atoms 
are calculated using Green function formalism. It is shown that the account of
 lattice relaxation and on-site Coulomb interaction are important 
for accurate description of magnetic properties of the investigated nanosystems.
 We predict a weak ferromagnetism phenomenon in the Mn antiferromagnetic nanochains on the CuN surface. The value of a net magnetic moment and direction of spin canting are calculated. 
We show that some experimental features may be explained using anisotropic exchange interactions.
\end{abstract}

\maketitle

\section{Introduction}
The investigation of magnetic nanomaterials is important part of nanoscience and exerts influence 
on progress in different sectors of technology (such as medicine devices for therapy and diagnostics, \cite{Zhang, Neuberger} magnetic data storage systems,\cite{Bajalan} etc). In the presence of intrinsic magnetic moment scientists can change a physical properties of nanomaterials
by applying external magnetic field. The technological applications of magnetic nanomaterials 
should base on an accurate control of the coupling between individual spins.

Recently, Hirjibehedin {\it et al.} \cite{IBM} have reported the fabrication of Mn nanoparticles in the form of linear
chains that display truly collective quantum behaviour. Using local spin-excitation spectroscopy technique, based on inelastic scanning tunneling microscopy
(STM), they were able to show how the quantum properties of this system depend on the number of atoms involved. They demonstrated an innovative method to measure and control these magnetic interactions. 
The experimental spectrum was analyzed using the simplest form of 
Heisenberg model
with exchange interaction only between nearest neighbors. However, there are a number of experimental results which cannot be explained by authors of Ref.\onlinecite{IBM} using the Heisenberg model:
(i) zero-field splitting which grows in energy with increasing chain length,
(ii) the different zero-field energy of the $m=\pm 1$ and $m=0$ excited states and 
(iii) asymmetry of the spectra with respect to voltage polarity.

Jones and Lin \cite{Jones} have applied GGA+U approach to describe the electronic and magnetic structures
of single and pair of Mn atoms on the CuN(100) surface. The performed spin-density analysis shows that Mn atoms on such surface 
preserve their atomic spins $S=\frac{5}{2}$. This result agrees with STM measurement.\cite{IBM} 
Electron-density change and surface relaxation due to Mn atoms are also analyzed in Ref.\onlinecite{Jones}.

The combination of experimental STM approach and theoretical {\it ab-initio} methods \cite{IBM2} has been used in order to describe the large magnetic anisotropy of individual Mn and Fe atoms on the CuN surface. The authors of the paper \cite{IBM2} have provided the detailed phenomenological picture of magnetic anisotropy and concluded that in case of manganese system the easy axis is oriented out-of-plane.   

In this paper we show that the local distortion of the system results in a superexchange interaction between Mn atoms 
through N atoms.
Isotropic exchange interactions are calculated using Green functions approach and total energies difference method. 
Using full diagonalization of Heisenberg Hamiltonian with calculated 
isotropic exchange integrals we estimate the energies of first 
magnetic excitations. The results are in good agreement with experimental data.

In the previous theoretical investigations a non-collinear magnetic ground state for
nanostructures on non-magnetic surfaces \cite{Gotsis,Eriksson} and magnetic \cite{Lounis} were proposed. For instance, the results
for the non-collinear triangular compact trimer of Cr on the Au(111) \cite{Gotsis} predict that the angle between
each pair of moments equals to 120$^{\circ}$ and the total spin moment is zero. Therefore, the non-collinearity is result of frustration of magnetic interactions.

In the paper \cite{Eriksson} authors have investigated different geometries of Fe, Mn and Cr atoms on the Cu(111) surface. The Fe clusters were found to be ferromagnetically ordered. Whereas for the Mn and Cr clusters an antiferromagnetic exchange interactions between nearest neighbours have been found. The antiferromagnetic couplings produce either collinear or non-collinear magnetic structures due to frustration of cluster geometry. 

An interesting results for the trimer and tetramer configurations of Mn and Cr atoms on the magnetic Ni(111) surface were obtained in the paper. \cite{Lounis} One should stressed that there are two types of a magnetic frustration: (i) frustration within adcluster and (ii) frustration arising from competing magnetic interactions between the adclusters and the surface atoms.  
 
Thus, one can conclude that the only known source for spin non-collinearity of magnetic clusters on nonmagnetic 3d surface is geometrical frustration which results in magnetic frustration. In this paper we propose a new source of spin non-collinearity for nanosystems on a surface. According to our calculations a local distortion between Mn atoms in the nanochain results in the strong Dzyaloshinskii-Moriya (DM) interaction. An important role plays the displacement of N atom from the surface. Based on first-principle calculations of the Dzyaloshinskii-Moriya interactions between magnetic moments we point out that the Mn nanochains on the CuN demonstrate a weak ferromagnetism phenomenon. We have estimated the value of a net magnetic moment and direction of spin canting. These results are also confirmed by direct LDA+U+SO calculations.
 
The paper is organized as follows. In Section II we describe the methods of the investigation. 
In Section III A and III B we present the results of LSDA and LDA+U calculations, respectively. 
The analysis and comparison of obtained exchange interactions with experimental data are presented in Section III C. Section IV is devoted to the analysis of zero-field energy splitting observed in STM experiment and in section V we briefly summarize our results.   

\section{Methods of investigation}

\subsection{DFT calculation details}
We have used two complementary approaches for investigations of an electronic and magnetic properties of Mn nanochains on the CuN surface. 

(i) First-principles total-energy and force calculations were carried out using the projected augmented-wave (PAW) method \cite{PAW} as implemented 
in the Vienna \emph{ab initio} simulation package (VASP). \cite{VASP,kresse} 
Exchange and 
correlation effects have been taken into account using LSDA and 
LDA+U \cite{Anisimov} approaches.
In all cases under investigation we used an energy cutoff of 400 eV in the
plane-wave basis construction and the energy convergence criteria of
$10^{-4}$ eV. The atomic positions of considered systems were relaxed with
residual forces less than 0.01 eV/\AA. For the Brillouin zone integration, 
a (4x4x1) Monkhorst-Pack mesh \cite{Monkhorst} and Gaussian-smearing 
approach with $\sigma=0.2$ eV were used.

To simulate structure of the unit cell we have used a supercell approach.
Structure of the supercell has consisted of two-layer (2{ }$\times${ }(n+1)) Cu(100) 
surface, N atoms embedded into upper Cu-layer, $\textrm{Mn}_{n}$-chains placed on
the top of the CuN-surface and vacuum region of 10 \AA. Lattice constant
for Cu was chosen to be 3.63 \AA, which gives a minimal value of the total 
energy in calculation of the bulk fcc Cu.
Lower layer of Cu has been fixed under relaxation.

(ii) We have also used the Tight Binding Linear-Mufin-Tin-Orbital Atomic Sphere 
Approximation (LMTO) method \cite{OKA} in terms of the conventional local density approximation taking into account the on-site 
Coulomb interaction LDA+U and spin-orbit coupling LDA+U+SO. \cite{Sol,shorikov} 
In this type of calculations we have used the relaxed structures 
obtained by PAW approach. The radii of atomic spheres were r(Mn)=1.137 \AA, r(Cu)=1.322 \AA ~and r(N)=0.793 \AA. In order to fill the empty space of the unit cell required number of empty spheres were added.       
\subsection{Spin Hamiltonian approach}

The main aim of our investigation is first-principle determination of parameters of the following spin Hamiltonian:
\begin{eqnarray}
H = H_{Heis} + H_{DM}, 
\end{eqnarray}
where Heisenberg \cite{Heisenberg} energy term is
\begin{eqnarray}
\label{heis}
H_{Heis} = \sum_{i < j} J_{ij} \vec S_{i} \vec S_{j}, 
\end{eqnarray}
and Dzyaloshinskii-Moriya \cite{Moriya} energy term is
\begin{eqnarray}
H_{DM} = \sum_{i < j} \vec D_{ij} [\vec S_{i} \times \vec S_{j}]. 
\end{eqnarray}

In order to calculate the isotropic exchange interactions J$_{ij}$, in Eq.(2) between magnetic moments of Mn atoms we have used two different approaches. (i) The total energy difference method on the basis of PAW results. The main idea of this method is that the isotropic exchange interaction defines
through the energy differences between different magnetic configurations. 
For instance, in case of Mn-dimer, spin Hamiltonian of the system can be written in the following form: 
\begin{equation}
H= J  \vec S_{1} \cdot \vec S_{2}.
\end{equation}
The corresponding total energies of the ferromagnetic and antiferromagnetic configurations
of two classical spins are given by
\begin{equation}
E_{FM}=  J S^{2}
\end{equation}
and
\begin{equation}
E_{AFM}= - J S^{2}.
\end{equation}
Therefore, the exchange interaction $J$ is expressed in the following form:
\begin{equation}
J = \frac{E_{FM}-E_{AFM}}{2S^{2}}.
\end{equation}
(ii) From the other hand, based on LMTO results one can calculate the isotropic exchange integrals and Dzyaloshinskii-Moriya interactions between magnetic moments of Mn atoms (S=$\frac{5}{2}$) using the local force theorem and Green functions formalism \cite{Lichtenstein, Solovyev, Mazurenko}
\begin{eqnarray}
J_{ij} = \frac{1}{2\pi S^2} \int_{-\infty}^{E_{F}} d\epsilon \, {\rm Im} \quad \quad \quad \quad \quad \nonumber \\
 \sum_{\substack {m, m' \\ m'', m'''}} (\Delta^{mm'}_{i} \,
G_{ij \, \downarrow}^{m'm''} \, \Delta^{m'' m'''}_{j} \, G_{ji \, \uparrow}^{m''' m}), 
\end{eqnarray}
where $m$ ($m^{'}$, $m^{''}$, $m^{'''}$) is magnetic quantum number and the on-site potential $\Delta^{mm'}_{i}=H^{m m'}_{ii \, \uparrow} - H^{m m'}_{ii \, \downarrow}$.
The Green function is calculated in the following way:
\begin{eqnarray}
G^{mm'}_{ij \sigma}(\epsilon) \, = \, \sum_{\mathbf{k},\, n} \frac{c^{mn}_{i \sigma} \, (\mathbf{k}) \, c^{m'n \, *}_{j \sigma} \,
(\mathbf{k})}{\epsilon-E^{n}_{\sigma}(\mathbf{k})}.
\end{eqnarray}
Here $c^{mn}_{i\sigma}$ is a component of the {\it n}-th eigenstate, E$_{\sigma}^{n}$ is the corresponding eigenvalue and $\bf{k}$ is quasimomentum in the first
Brillouin Zone.

In turn the Dzyaloshinskii-Moriya interaction, Eq.(3) can be calculated through the account
of spin-orbit coupling in the second variation of total energy of the system over the small 
deviations of magnetic moments from the collinear ground state: \cite{Mazurenko,SolovyevDM}
\begin{eqnarray}
D^{z}_{ij} = - \frac{1}{2 \pi S^2} \, Re \int_{-\infty}^{E_{F}} d \epsilon \, \sum_{k} 
\nonumber \\
\times (\Delta_{i} G_{ik}^{\downarrow} H^{so}_{k \, \downarrow \downarrow} G_{kj}^{\downarrow} \Delta
_{j} G_{ji}^{\uparrow} -
\Delta_{i} G_{ik}^{\uparrow} H^{so}_{k \, \uparrow \uparrow} G_{kj}^{\uparrow} \Delta_{j} G_{ji}^{\downarrow} 
\nonumber \\
+ \Delta_{i} G_{ij}^{\downarrow} \Delta_{j} G_{jk}^{\uparrow} H^{so}_{k \, \uparrow \uparrow} G_{ki}^{\uparrow} -
\Delta_{i} G_{ij}^{\uparrow} \Delta_{j} G_{jk}^{\downarrow} H^{so}_{k \, \downarrow \downarrow} G_{ki}^{\downarrow}),
\end{eqnarray}
where $H^{so}_{k}=\lambda_{k} \vec L \vec S$ and $\lambda_{k}$ is spin-orbit coupling constant for site $k$.
Here we present only $z$ component of Dzyaloshinskii-Moriya vector. $x$ and $y$ components can be obtained from the $z$ ones by rotation of the coordinate system. 

\section{Results} 
\subsection{LSDA results}
\begin{figure}[!ht]
\includegraphics[width=0.45\textwidth, angle=0]{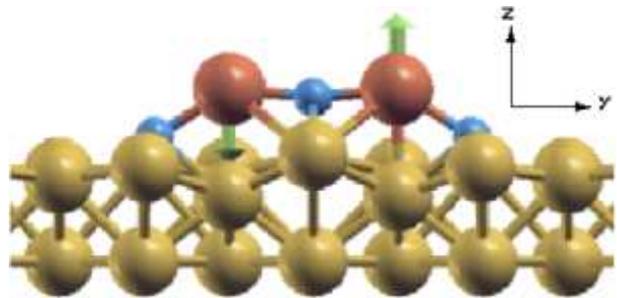}
\caption{The projection on yz-plane of relaxed structure of Mn-dimer supported on the CuN obtained using LSDA approach. Yellow, blue and red spheres correspond to Cu, N and Mn atoms, respectively. Green
arrows correspond to direction of atomic magnetic moments. \\}
\label{dimer_lsda}
\end{figure}

We have performed LSDA calculation of antiferromagnetic Mn-dimer supported on the CuN surface. Fig.\ref{dimer_lsda} shows the relaxed structure of the Mn-dimer obtained within LSDA method using PAW approach. The information about structure of relaxed system is presented in Table I.
One can see that Mn-N-Mn bond angle of 171$^{\circ}$ is close to 180$^{\circ}$ and corresponds to the maximum of superexchange interaction between 3d atoms.

\begin{figure}[]
\includegraphics[width=0.47\textwidth,angle=0]{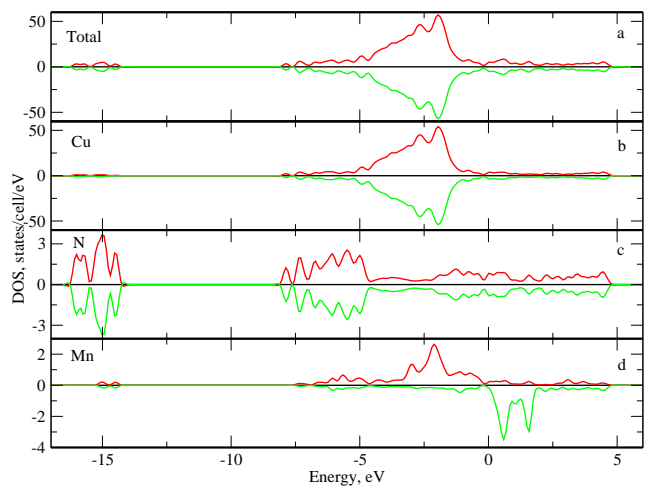}
\vspace{0.2cm}
\caption{Total and partial density of states obtained using LSDA calculations. (a) Total DOS, (b)
Projected DOS onto Cu atoms, (c) Projected DOS onto N atoms, (d) Projected DOS
onto Mn atom. Fermi level corresponds to 0 eV.}
\label{dos_lsda}
\end{figure}

The calculated total and partial densities of states obtained using PAW approach are presented in Fig. \ref{dos_lsda}.
The valence band contains low and high energy parts which are 
separated by 6 eV. The low energy states located around -15 eV are mainly N 
ones. In the region from -7 eV to 5 eV all states are highly mixed. 

The calculated values of magnetic moments of Mn atoms within LMTO and PAW approaches are 3.70 $\mu_{B}$ and 
3.35 $\mu_B$, respectively. These values are 
smaller than experimentally observed spin $\frac{5}{2}$. 
Moreover, the calculated exchange parameter $J_{ij}$ within Green functions approach, Eq.(8)  
is 20.4 meV, whereas total energies difference method value, Eq.(7) is 24.8 meV. These values at least three times larger than that experimentally observed. 
Thus, one can see that the main electronic and magnetic properties of the 
investigated nanosystem cannot be correctly reproduced within the LSDA approach.

\subsection{LDA+U results}

\begin{figure}[!ht]
\includegraphics[width=0.33\textwidth]{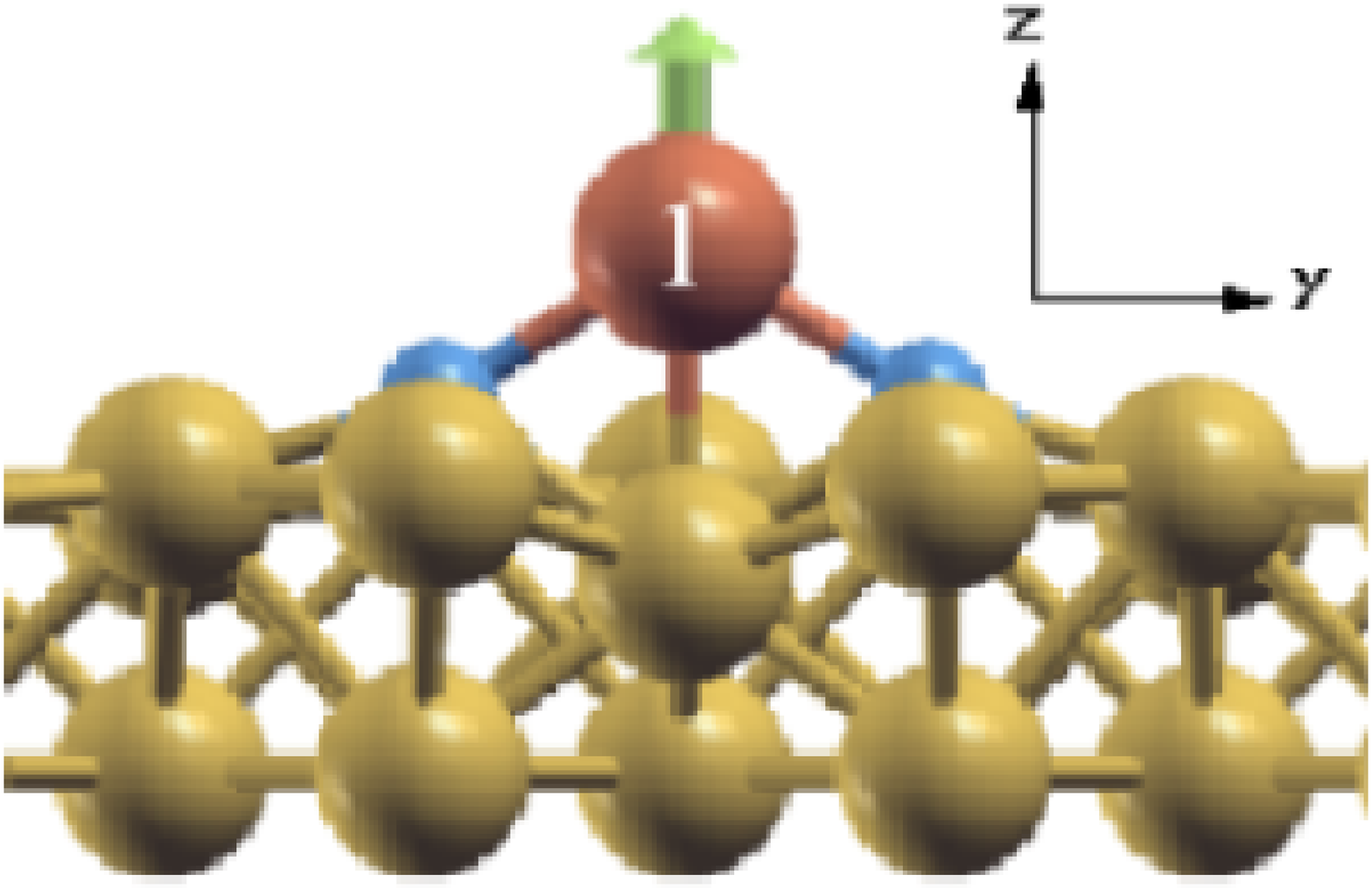}
\vspace{0.1cm}
\includegraphics[width=0.33\textwidth]{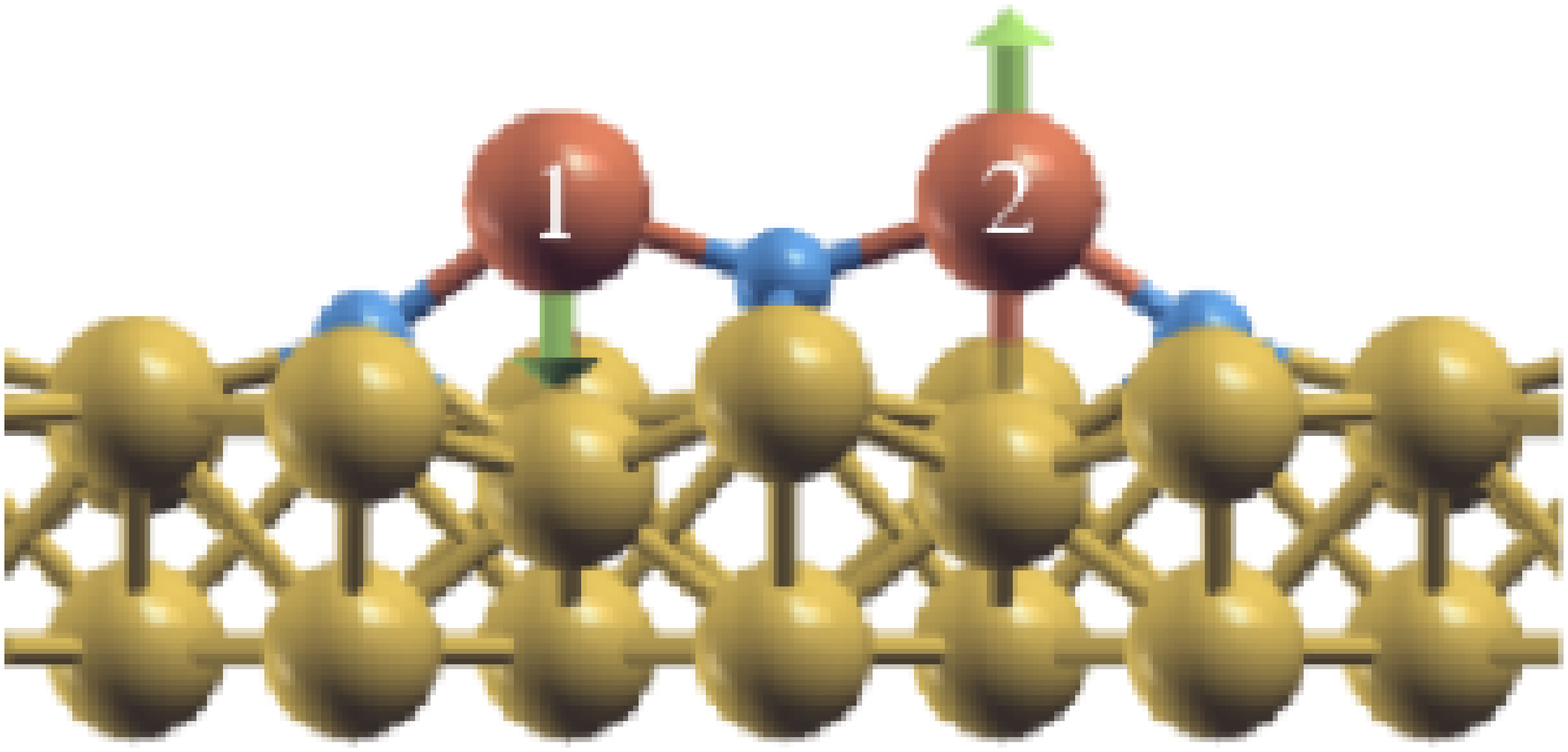}
\vspace{0.1cm}
\includegraphics[width=0.33\textwidth]{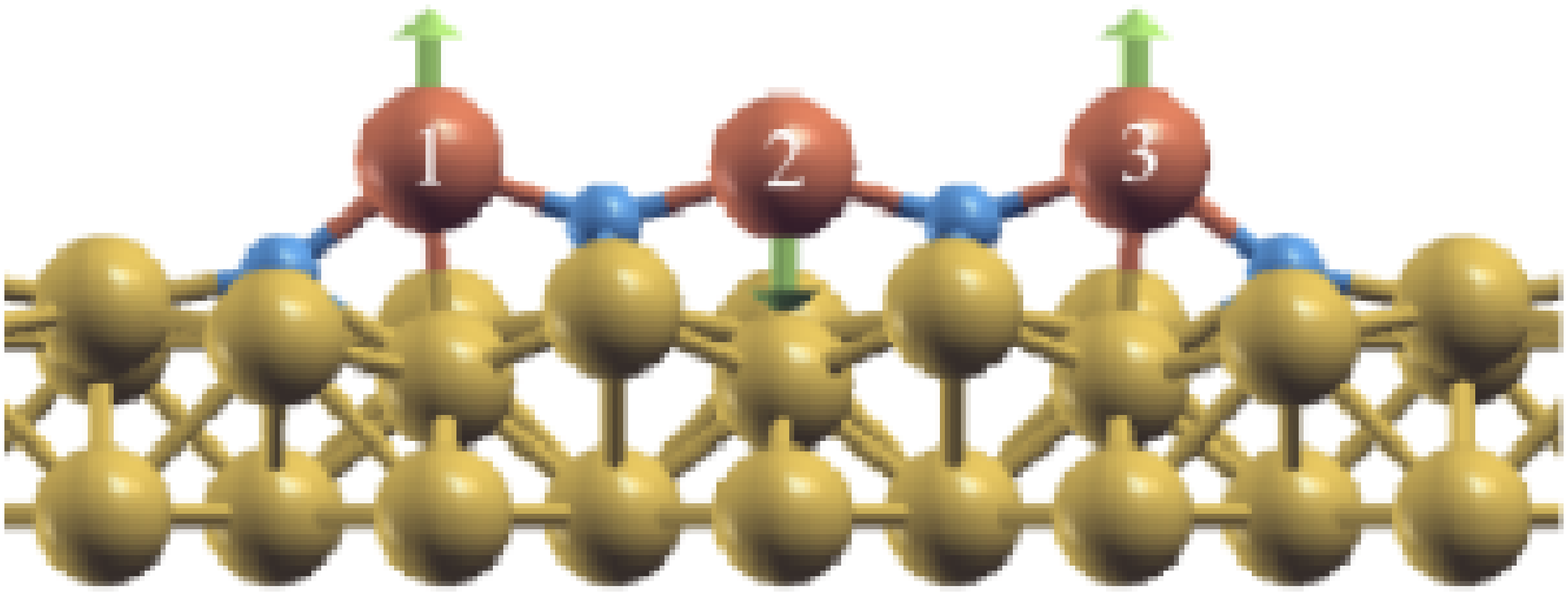}
\vspace{0.1cm}
\includegraphics[width=0.33\textwidth]{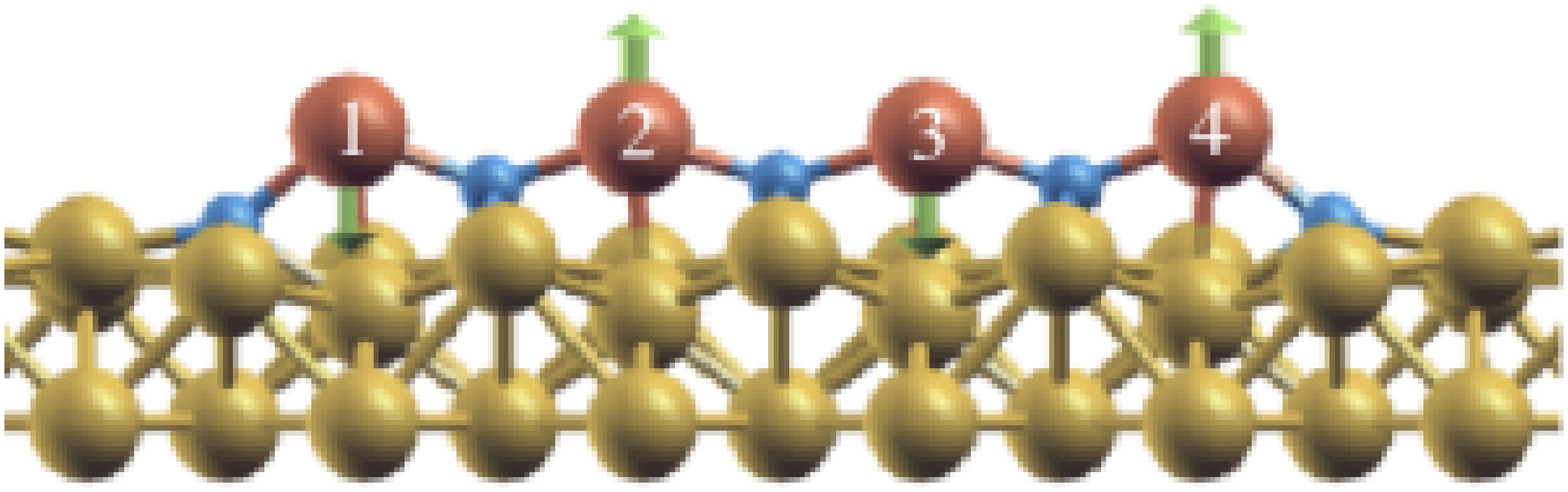}
\caption{The yz projections of relaxed structures of Mn chains of CuN surface for n=1 $\div$ 4. Yellow, 
blue and red spheres correspond to Cu, N and Mn atoms, respectively. Green
arrows correspond to direction of atomic magnetic moments of manganese.}
\label{structures}
\end{figure}

The results of the previous section have demonstrated drawbacks of the LSDA approach to describe
the magnetic properties of the Mn dimer on the CuN surface. It is well known
problem of local density approximation in respect to transition metal 
compounds.
To overcome this problem we have used 
the LDA+U approach with on-site Coulomb and on-site Hunds interaction 
parameters of $U$=6.0 eV and $J_{H}$=0.9 eV, respectively.
These values are in good agreement with recent first-principle estimations performed in the work. \cite{IBM2}

The relaxed structures of the Mn-chains of different lengths (n=1 $\div$ 4) obtained within PAW calculations are shown in Fig.\ref{structures}. The structural information is presented in Table \ref{struct}. The obtained structures 
have some interesting geometrical features. Let us analyze the difference between the clean CuN-substrate and the substrate with
Mn adatoms on the top. Presence of Mn atoms causes some rearrangement of
upper layer atoms of CuN-substrate. Generally, this rearrangement concerns the 
N atoms. In contrast to clean-CuN surface, the N atoms of the system with Mn nanochains 
are significantly shifted from the first layer plane. This fact agrees with results of recent GGA calculations. \cite{Jones}
The calculated angle of Mn-N-Mn bond within LDA+U approximation equals to $143^\circ$. The N atoms at the edges of chains also have some displacement
from the plane in z-direction, but to a smaller extent than N atoms situated inside the chain.  

\begin{figure}[]
\includegraphics[width=0.47\textwidth,angle=0]{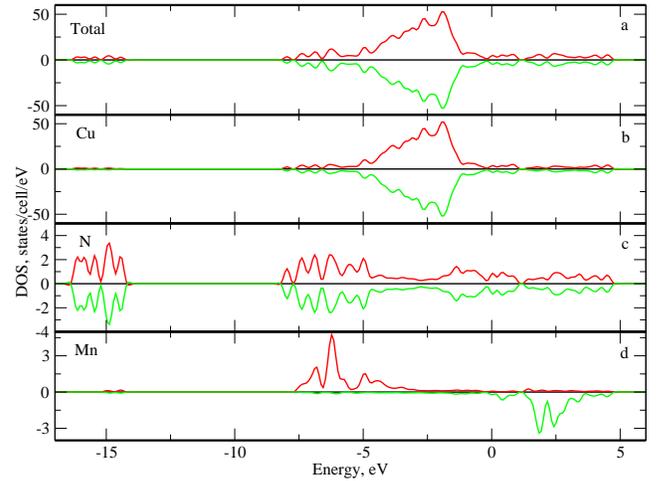}
\vspace{0.2cm}
\caption{Total and partial density of states of Mn-dimer on CuN surface obtained using LDA+U calculations. (a) Total DOS, (b) Projected DOS onto Cu atoms, (c) Projected DOS onto N atoms, (d) Projected DOS 
onto Mn atom. Fermi level corresponds to 0 eV.}
\label{dos_ldapu2}
\end{figure}

From a geometrical point of view the important difference between results of LSDA and LDA+U 
approaches is the angle of Mn-N-Mn bond.
Let us analyze this fact on the level of hopping integral.
For simplicity, we assume that there is the only strong hopping between orbitals of two 3d atoms. 
The hopping integral $t_{ij}$ is proportional to $\cos \alpha$, where $\alpha$ 
is angle of metal-ligand-metal bond.
Therefore, within LDA+U approach the hopping integral is strongly suppressed
due to local Coulomb correlations.
In turn the isotropic exchange interaction between Mn atoms in the atomic limit of Hubbard model can be expressed 
as $J_{ij}= \frac{4 t_{ij}^{2}}{U}$, here $U$ is on-site Coulomb integral. 
It is clear that on-site Coulomb interaction is another source of suppression of the isotropic exchange interaction in LDA+U in comparison with LSDA approach. 
\begin{figure}[!ht]
\includegraphics[width=0.47\textwidth,angle=0]{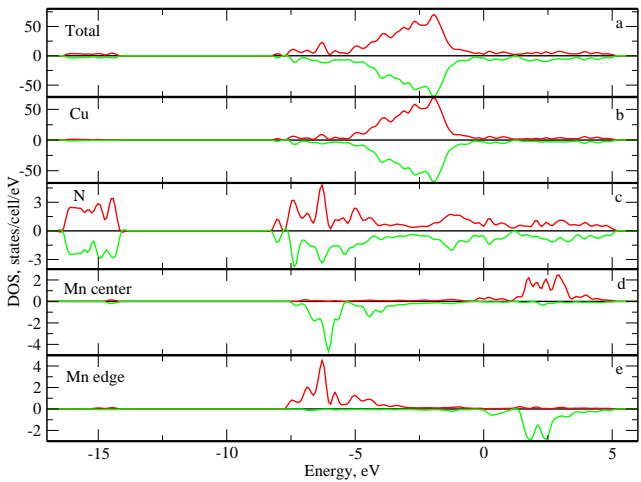}
\vspace{0.3cm}
\caption{Total and partial density of states of Mn-trimer chain on CuN surface  
obtained using LDA+U calculations. (a) Total DOS, (b) Projected DOS onto Cu atoms, (c) Projected DOS onto N atoms, (d) Projected DOS 
onto central Mn atom, (e) Projected DOS onto edge Mn atom.}
\label{dos_ldapu3}
\end{figure}

The total and partial densities of states of the dimer and trimer systems obtained using LDA+U 
approximation are presented in Fig.\ref{dos_ldapu2} and \ref{dos_ldapu3}, respectively.
The calculated value of magnetic moment are listed in
Table \ref{values}. One can see that the values of magnetic moments of
Mn atoms within the chains varies from 4.34 $\mu_B$ (middle atoms) to 
4.47 $\mu_B$ (edge atoms) and now is much closer to experimental values of
 $S=\frac{5}{2}$. In the case of LMTO results this difference is much smaller.

\begin{table}[!ht]
\centering
\caption[Bset]{Structural information about alignment Mn and N atoms on
the Cu(100) surface as result of relaxation in cases of LSDA and LDA+U 
approaches. The values are distance (in \AA) between Mn atoms, $z$-coordinates of Mn atoms and N atoms at the center and edge of chain, respectively. $\alpha$ is angle of Mn-N-Mn bond. Zero level of z coordinate corresponds to the lower layer of Cu atoms. All values are given in \AA.}
\label{struct}
\begin{tabular}{ccccccc}
  \hline
  \hline
&   &  $d_{Mn-Mn}$ & $z_{Mn}$ & $z_{N}(center)$ & $z_{N}(edge)$ & $\alpha$\\
  \hline
&  LSDA     &  3.58   & 3.66 & 3.49 & 2.66 & 171$^{\circ}$ \\
&  LDA+U   &  3.78   & 3.75 & 3.11 & 2.59 & 143$^{\circ}$ \\
  \hline
  \hline
\end{tabular}
\end{table}

\begin{table}[!ht]
\centering
\caption[Bset]{Values of magnetic moments of Mn atoms (in $\mu_B$) calculated using LMTO (PAW) method.}
\label{values}
\begin{tabular}{cccc}
  \hline
  \hline
& n  &  M$_{edge}$ & M$_{center}$  \\
  \hline
&  1   &   -   & 4.46 (4.48)    \\
&  2   &  4.45 (4.44)  &     -        \\
&  3   &  4.47 (4.47)   & 4.45 (4.34) \\
&  4   &  4.47 (4.47)   & 4.40 (4.35)    \\
  \hline
  \hline
\end{tabular}
\end{table}

\subsection{Isotropic exchange interaction}
The next step of our investigation is determination of Heisenberg exchange interaction parameters in Eq.(2).
The magnetic couplings between Mn atoms calculated using Green functions, Eq.(8) and the total energy difference method, Eq.(7) are presented in Table \ref{exch}. One can see that the calculated values of the dimer interaction are in good
agreement with experimental value of 6.4 meV. The value of isotropic
exchange integral between nearest Mn atoms in trimer is smaller than that in dimer system. There is also small ferromagnetic coupling 
between edge Mn atoms. 

In the case of quatromer system there is the difference between exchange
integrals J$_{12}$ and J$_{23}$. The coupling at the center
of the chain is smaller than the coupling at the edge. There is interesting
dependence of nearest-neighbour exchange interaction according to the chain length. This
tendency probably corresponds to oscillations of exchange interaction 
parameter depending on chain length. It is important to
investigate the mechanism of such strong oscillations (see Mn$_3$ results
in Table \ref{exch}) of nearest
neighbour exchange interactions in Mn$_n$-nanochains. Such analysis is left
for future investigation.

Using full diagonalization procedure of ALPS library \cite{ALPS1,ALPS2} we have calculated the spin excitation spectra of investigated systems. 
The energies of first excited states are presented in Table \ref{exitate}. Despite of the fact that the $J_{12}$ of trimer has smaller value than in dimer case, weak ferromagnetic $J_{13}$ interaction between edge atoms compensates this difference and gives us opportunity to reproduce experimentally observed excitation energy with reasonable accuracy. 

\begin{table}[!ht]
\centering
\caption[Bset]{Values of exchange interactions J$_{ij}$ (in meV) between 
magnetic moments of Mn atoms calculated using Green function method (TB-LMTO-ASA) method. Values obtained using total energies difference method (PAW) approach are given in parenthesis.}
\label{exch}
\begin{tabular}{ccccccc}
  \hline
  \hline
 n  & J$_{12}$ & J$_{13}$  & J$_{23}$  & J$_{34}$  &  J$_{24}$   \\
  \hline 
  2   &   7.0 (6.0)  &      -      &    -    &   -   &    -         \\
  3   &   4.0 (4.2)  &  -0.09 (-0.09)  &    4.0 (4.2)    &   -   &  -       \\
  4   &   5.6 (5.2)      &   -0.07 (-0.04)      &  2.4 (4.1)    &  5.6 (5.2)     &  -0.07 (-0.04)        \\
  \hline
  \hline
\end{tabular}
\end{table}

For quatromer system our results are in excellent agreement with experimental spectrum.
\begin{table}[!ht]
\centering
\caption [Bset]{Energies of first excited states (in meV) of Heisenberg model obtained using ALPS code for the chain systems of different length.} 
\label{exitate}
\begin {tabular}{ccccc}
  \hline
  \hline
 $n$  & $E^{exp}$ & $E^{calc}_{LMTO}$   &  $E^{calc}_{PAW}$     \\
  \hline
  2   &    6.4      &     7.0       &    6.0           \\
  3   &    16.0      &     10.5    &   10.9      \\
  4   &    2.9      &     3.0      &    2.6          \\
  \hline
  \hline
\end {tabular}
\end {table}

\subsection{Dzyaloshinskii-Moriya interaction}
\begin{figure}[!ht]
\includegraphics[width=0.37\textwidth,angle=0]{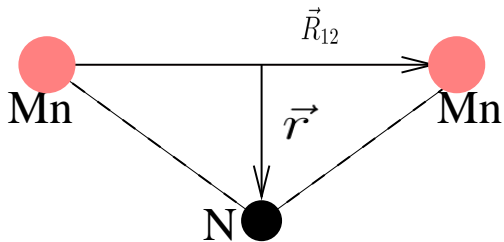}
\vspace{0.3cm}
\caption{Schematic representation of displacement of the ligand atom.}
\label{shift}
\end{figure}

From the crystal symmetry point of view there is no inversion center at the point bisecting the straight line between Mn atoms of investigated nanosystems. Therefore, in according with Moriya's rules \cite{Moriya} a DM coupling exists. First, let us perform a simple geometrical analysis of the symmetry of Dzyaloshinskii-Moriya vector. There are two sources of inversion symmetry breaking in the investigated systems.
(i) The first one is the substrate surface. Based on the fact that previous investigations of metallic nanochains on nonmagnetic 3d surface has no sign of non-collinearity, one can conclude that surface gives 
negligible small contribution to anisotropic exchange interaction. (ii) More importantly, the second source of inversion symmetry breaking is vertical displacement of N atom. The Dzyaloshinskii-Moriya vector, $\vec D_{12}$ between Mn atoms is proportional to 
$[\vec r \times \vec R_{12}]$,\cite{Mostovoy} where $\vec R_{12}$ is a unit vector along the line connecting 
the magnetic ions and $\vec r$ is the shift of the ligand atom from this line (Fig. \ref{shift}). One can see that in our case $\vec r$ and $\vec R_{12}$ have z and y components, respectively. Therefore, the direction of Dzyaloshinskii-Moriya vector is x axis. 

The calculated 
anisotropic exchange interactions, Eq.(10) are presented in Table \ref{dzyaloshinskii}. For all systems under consideration 
the Dzyaloshinskii-Moriya vector lies along x axis (perpendicular to the Mn chain and parallel to the CuN surface).  
Therefore, if all spins lie in the yz plane, the canting exists and there
is weak ferromagnetism in the system.
The ratio between Dzyaloshinskii-Moriya and isotropic exchange interactions,
 $\frac{|\vec D_{12}|}{J_{12}}$=0.002 is the same order of magnitude as in case of 
 well known antiferromagnets Fe$_{2}$O$_{3}$ and La$_{2}$CuO$_{4}$ with weak ferromagnetism.  
\begin{table}[!ht]
\centering
\caption [Bset]{Values of x component of  Dzyaloshinskii-Moriya interactions $\vec D_{ij}$ (in meV) between 
magnetic moments of Mn atoms calculated using Green functions method (Eq.(10)) within LDA+U approach.}
\label{dzyaloshinskii}
\begin {tabular}{cccc}
  \hline
  \hline
   $n$    & $D^{x}_{12}$ & $D^{x}_{13}$  & $D^{x}_{23}$\\
  \hline
    2   &   0.014   &      -    &    -    \\
  3   &     0.018   &   0.000  &   0.018   \\
  4   &     0.024   &   -0.006  &   0.030   \\
  \hline
  \hline
\end {tabular}
\end {table}

\begin{figure}[!ht]
\includegraphics[width=0.1\textwidth,angle=0]{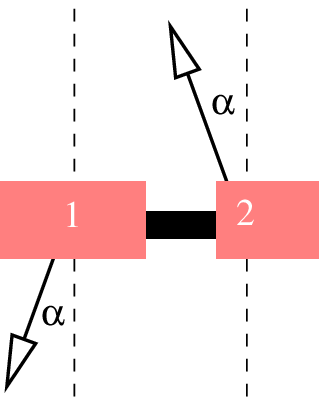}\\
\includegraphics[width=0.15\textwidth,angle=0]{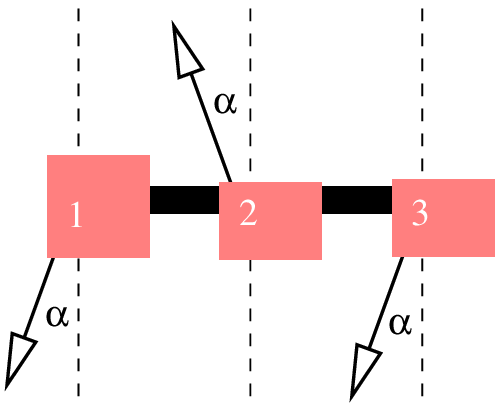}\\
\includegraphics[width=0.25\textwidth,angle=0]{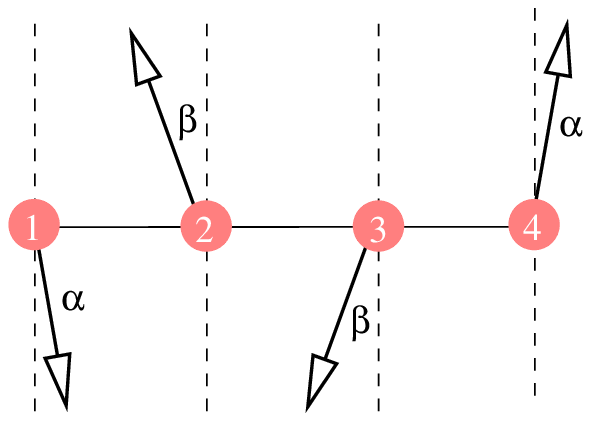}
\vspace{0.2cm}
\caption{Schematic representation of spin canting of dimer, trimer and quatromer systems.}
\label{spin_canting}
\end{figure}

We have minimized the classical spin Hamiltonian (Eq.(1)) with first-principles exchange parameters in respect to the angle between 
different spins in the chain. On this basis we have defined the values of canting angles and net magnetic moments of the Mn nanochains. 
These results are presented in Table \ref{canting} and Fig. \ref{spin_canting}. 
One can see that the net magnetic moment increases with length of the nanochain.

In order to test the reliability of the weak ferromagnetism results we have performed the LDA+U+SO calculations. 
For dimer system the magnetic ground state is non-collinear and spins are along z axis with canting of 1.0$^{\circ}$ for LMTO and 1.6$^{\circ}$ for PAW. 
These results are in good agreement with previous GGA calculations \cite{IBM2} where the easy axis of a system with single Mn atom on the CuN surface is z axis. 
The obtained values of canting angles are about one order larger than those 
obtained in our LDA+U calculations (Table \ref{canting}). This observation can be addressed to underestimation of Dzyaloshinskii-Moriya interaction calculated by Green functions method.\cite{Mazurenko}  
Obtained canting angles correspond to the following Dzyaloshinskii-Moriya interactions $D^x_{ij} = J_{ij} tg (\pi - 2 \alpha)$ = 0.24 meV (in LMTO) and $D^x_{ij}$ = 0.34 meV (in PAW). One can consider these results
as manifestation of strong Dzyaloshinskii-Moriya interaction in investigated nanosystems.
  
\begin{table}[!ht]
\centering
\caption [Bset]{Values of canting angles and full weak ferromagnetic moments of different Mn nanochains (in $\mu_{B}$) obtained using minimization of model spin Hamiltonian.}
\label{canting}
\begin {tabular}{ccc}
  \hline
  \hline
   $n$   & angle                                  & $M_{model}$  \\
  \hline
  2      &  $\alpha$=0.057$^{\circ}$                     &   0.009           \\
  3      &  $\alpha$=0.164$^{\circ}$                            &   0.038           \\
  4      &  $\alpha$=0.173$^{\circ}$ $\beta$=0.466$^{\circ}$    &   0.045              \\
  \hline
  \hline
\end {tabular}
\end {table}

Clearly, the ultimate test of our results will to compare them with experiment. In the next section 
we will show that some experimentally observed features \cite{IBM} can be explained using the anisotropic exchange interaction.

\section{Zero-field splitting}  
In order to explain experimentally observed different zero-field energy of $m = \pm 1$ and $m=0$ excited states \cite{IBM} we use the quantum spin Hamiltonian with Dzyaloshinskii-Moriya interaction:
\begin{eqnarray}
\hat H = J \hat {\vec S}_{1} \hat {\vec S}_{2} + \vec D_{12} [\hat {\vec S}_1 \times \hat {\vec S}_2].
\end{eqnarray}

For simplicity, let us consider the case of S=$\frac{1}{2}$ and $\vec D_{12} = (D^x_{12}; 0; 0)$. One can rewrite Eq.(11) in the following form:
\begin{eqnarray}
\hat H = J (\hat S_1^x \hat S_2^x + \hat S_1^y \hat S_2^y + \hat S_1^z \hat S_2^z ) + D^x_{12} (\hat S_1^y \hat S_2^z - \hat S_1^z \hat S_2^y).
\end{eqnarray}
The basis functions for this Hamiltonian can be written as follows,
\begin{eqnarray}
| \uparrow, \uparrow > \quad \quad | \downarrow, \downarrow > \quad \quad | \uparrow, \downarrow> \quad \quad | \downarrow, \uparrow >.  
\end{eqnarray}
Using well known rules for spin operators 
\begin{eqnarray}
\hat S^x | \uparrow > = \frac{1}{2} | \downarrow > \quad \quad \hat S^y | \uparrow > = \frac{i}{2} | \downarrow > \quad \quad \hat S^z | \uparrow > = \frac{1}{2} | \uparrow > \nonumber \\
\hat S^x | \downarrow > = \frac{1}{2} | \uparrow > \quad \quad \hat S^y | \downarrow > = - \frac{i}{2} | \uparrow > \quad \quad \hat S^z | \downarrow > = -\frac{1}{2} | \downarrow > 
\nonumber
\end{eqnarray}
one can define the matrix elements of this Hamiltonian presented in Table \ref{matrix}.
\begin{table}[!ht]
\centering
\caption [Bset]{Matrix elements of the Heisenberg Hamiltonian Eq.(12)}
\label{matrix}
\begin {tabular}{ccccc}
  \hline
  \hline
                                & $|\uparrow, \uparrow >$ & $|\downarrow, \downarrow >$  & $|\uparrow, \downarrow >$   & $|\downarrow, \uparrow >$     \\
  \hline
  $|\uparrow, \uparrow >$       &  $\frac{J}{4}$    & 0            & $\frac{i D^x_{12}}{4}$   &  $-\frac{i D^x_{12}}{4}$    \\
  $|\downarrow, \downarrow >$   &  0              & $\frac{J}{4}$  & $-\frac{i D^x_{12}}{4}$   &  $\frac{i D^x_{12}}{4}$    \\
  $|\uparrow, \downarrow >$     & $-\frac{i D^x_{12}}{4}$     & $\frac{i D^x_{12}}{4}$  & $-\frac{J}{4}$   & $\frac{J}{2}$     \\
  $|\downarrow, \uparrow >$     & $\frac{i D^x_{12}}{4}$     & $-\frac{i D^x_{12}}{4}$  & $\frac{J}{2}$   & $-\frac{J}{4}$      \\
  \hline
  \hline
\end {tabular}
\end {table}
The eigenvalues of this matrix are the following:
\begin{eqnarray}
E_T^{\pm} = \frac{J}{4}, \nonumber \\
E_T^0= -\frac{J}{4} + \frac{\sqrt {J^2+4D^2}}{2},   \nonumber \\
E_S = -\frac{J}{4} - \frac{\sqrt {J^2+4D^2}}{2}. \nonumber
\end{eqnarray}
One can see that the energies of $m=0$ and $ m=\pm 1$ triplet states are different. 
Therefore, one can expect that anisotropic exchange interaction helps us explain similar difference in the experimental spectra for the Mn dimer on the CuN surface. 

Since in the case of S=5/2 the situation is more complicated, we have numerically calculated the excitation spectra 
using previously obtained isotropic and anisotropic exchange interactions by means of ALPS library. \cite{ALPS1,ALPS2} The final results are presented in Table \ref{en_diff}.
\begin{table}[!ht]
\centering                                                       
\caption [Bset]{The energy difference, $\Delta E$ of $m=0$ and $m= \pm 1$ excited states for different sets of calculated isotropic and anisotropic exchange interactions (in meV).}
\label{en_diff}
\begin {tabular}{cc}                                                                                     \hline
  set &   $\Delta E$ \\
  \hline
  LMTO (LDA+U): $J$ = 7.0, $D^x$= 0.02      & $< 10^{-4}$        \\
  LMTO (LDA+U+SO): $J$ = 7.0, $D^x$ = 0.24        &  0.02          \\
  PAW (LDA+U+SO): $J$ = 6.0, $D^x$ = 0.34     &    0.04               \\
  \hline                                                                                                  \hline
\end {tabular}
\end {table}
According to experiment the zero-field energies of excited states are $5.96 \pm 0.05$ meV 
 and $5.83 \pm 0.05$ meV for $m=0$ and $m = \pm 1$, respectively, and
correspond to $\Delta E=0.13 \pm 0.05$ meV. One can see that the 
theoretically estimated values of $\Delta E$ (Table \ref{en_diff}) give
the correct order of magnitude for experimental zero-field energy splitting.
We plan to investigate the effect of single-ion magnetic anisotropy energy
on this splitting.

\section{Conclusion}

We have performed 
first-principle investigations of electronic and magnetic structures of
the Mn nanochains supported on the CuN surface. Relaxation effects have taken into
consideration. The
calculated isotropic exchange integrals are in good agreement
with experimental data. We have also calculated the
anisotropic exchange interactions in the system and predicted the 
antiferromagnetic ground state with weak ferromagnetism. We stress that the 
main source of this phenomenon is local distortion which breaks the inversion symmetry between Mn atoms. It follows that the relaxation effects 
are important for the system under consideration.
The calculated values of canting angles are larger than those for classical antiferromagnetics with weak ferromagnetism,
Fe$_{2}$O$_{3}$ and La$_{2}$CuO$_{4}$. Using calculated anisotropic exchange interactions we have explained the experimentally observed different zero-field energies of $m=0$ and $m= \pm 1$ states. 

Based on our results one can expect the weak ferromagnetism phenomenon in the similar surface nanosystems. For instance, we found this spin-orbit coupling effect in Co nanochain on the Pt surface. Such work is in progress. 
 
\section{Acknowlegment}
We would like to thank F. Mila, M. Troyer, M. Sigrist, I.V. Solovyev and F. Lechermann for helpful discussions.
The hospitality of the Institute of Theoretical Physics of Hamburg University is gratefully acknowledged. This work is supported by DFG Grant  No. SFB 668-A3 (Germany), INTAS Young Scientist Fellowship Program Ref. Nr. 04-83-3230, 
Russian Foundation for Basic Research grant RFFI 07-02-00041, RFFI 06-02-81017, the grant program of President of Russian Federation
Nr. MK-1041.2007.2 and Intel Scholarship Grant.
The calculations were performed on the computer cluster of ``University Center of Parallel Computing'' of USTU-UPI 
and Gonzales cluster of ETH-Zurich.

\end{document}